\documentclass[manuscript,nonacm]{acmart}
\AtBeginDocument{%
  }

\usepackage{todonotes}
\usepackage{graphicx}
\usepackage{subcaption}
\begin{document}

\title{``We Are Tired of Explaining'': Communication Practice and AI Roleplay Training for Community Health Workers in Rural India}

\author{Neil K. R. Sehgal}
\email{neilsehgal99@gmail.com}
\affiliation{%
  \institution{University of Pennsylvania}
  \city{Philadelphia}
  \state{PA}
  \country{USA}
}

\author{Sunny Rai}
\affiliation{%
  \institution{University of Pennsylvania}
  \city{Philadelphia}
  \state{PA}
  \country{USA}
}

\author{Sai Preethi Matam}
\affiliation{%
  \institution{Mamata Academy of Medical Sciences}
  \city{Hyderabad}
  \country{India}
}

\author{Khushboo Gupta}
\affiliation{%
 \institution{Khushi Baby}
 \country{India}}

\author{Hamid Abdullah}
\affiliation{%
 \institution{Khushi Baby}
 \country{India}}

\author{Mohit Jain}
\affiliation{%
  \institution{Microsoft Research India}
 \country{India}
}

\author{Sharath Chandra Guntuku}
\affiliation{%
  \institution{University of Pennsylvania}
  \city{Philadelphia}
  \state{PA}
  \country{USA}
}

\renewcommand{\shortauthors}{Sehgal et al.}

\begin{abstract}
Community health workers (CHWs) in the Global South increasingly encounter AI-powered tools, yet the counseling work central to their role remains largely unsupported. We study communication practices among Accredited Social Health Activists (ASHAs) in rural Rajasthan, India, through simulated family-planning calls, semi-structured interviews, and an LLM chatbot roleplay design-probe with 20 participants. In calls, ASHAs often responded to social or material concerns by shifting to health-risk information, denying concerns, promising unspecified help, or listing medical solutions with limited explanation. A smaller set of responses instead engaged concerns, sought permission before involving family members, or left decisions with beneficiaries. We interpret these patterns through Motivational Interviewing, emphasizing restraint from correcting, persuading, or over-solving. Drawing across observed calls, interviews, and probe reactions, we derive design considerations for AI roleplay training: keep AI in a rehearsal role, provide descriptive rather than prescriptive feedback, and evaluate counseling process rather than agreement with prescribed responses.
\end{abstract}

\maketitle

\section{Introduction}

India's Accredited Social Health Activist (ASHA) program is one of the largest groups of community health workers (CHWs) in the world, with over 1 million ASHAs across the country \cite{CHAWLA2025100134}. ASHAs are local women recruited from the villages they serve, and are given 2-4 weeks of training to provide last-mile health services. Their tasks involve facilitating institutional deliveries, mother and child health tracking, distributing contraceptives and medications, and counseling families on family planning, nutrition, and child health \cite{CHAWLA2025100134}. Importantly, the work is not just clinical, but relational as well. For instance, their job involves counseling women and their families about services that they may be uncertain about including sterilization, contraception, vaccination, and antenatal care.

This work demands strong communication skills, and ASHAs operate inside a network of familial and social dynamics~\cite{shrivastava2016measuring}. For example, a woman wanting to pursue sterilization may be hesitant without her husband's or mother-in-law's support, or may not have help with household work during recovery~\cite{aruldas2017care}. A family with four daughters may resist contraception because they still hope for a son, a preference tied to old-age security, inheritance structures, or deeply held community norms~\cite{CHAWLA2025100134}. 

What counts as good counseling in this setting is therefore not obvious. Family planning scholarship increasingly defines quality through contraceptive autonomy, meaning whether a woman can make and act on an informed and voluntary decision, including a decision to delay or decline a method \cite{senderowicz2020contraceptive}. On this standard, uptake and family agreement are not by themselves evidence that a conversation went well, and an informed refusal is a legitimate outcome rather than a failure. We adopt this standard throughout, and Section~2.5 situates it in the history of family planning in India.

National ASHA training materials include both health-related content (e.g., family planning methods, vaccination plan, and pregnancy complications) and communication skills such as listening, negotiation, and handling difficult conversations \cite{nrhm2013asha,nhsrc_asha_induction_2021}. Despite this emphasis, multiple studies have identified shortcomings in ASHAs’ communication practices, including counseling on family planning, addressing vaccine hesitancy, and navigating sensitive conversations around COVID-19 \cite{shrivastava2016measuring,scott2022we,goel2019effectiveness}. These limitations can reduce ASHAs’ confidence, reducing the effectiveness of their home visits \cite{shrivastava2016measuring}. However, these studies rely on observer ratings, self-report, or downstream outcomes; none observes what ASHAs actually say when beneficiaries resist or challenge their advice, or raise social concerns, and how these conversations unfold turn by turn.

Recent HCI work has examined how ASHAs engage with AI tools and how digital health technologies fit into their workflows \cite{okolo2024if,okolo2021cannot,ramjee2025ashabot,wadhwa2025designing,ming2022invisible,solano2024explorable}. This literature documents the relational labor ASHAs perform, including trust-building, boundary-negotiation, and emotional work, and shows how little of it is visible to the AI systems designed to assist them \cite{ming2022invisible}. 
Counseling, as in responding to beneficiaries' concerns and helping them navigate difficult decisions, has received less attention.
Understanding this practice is important not only for characterizing ASHAs' work, but also for designing AI systems intended to support it.

To address this gap, we conducted a qualitative study with 20 ASHAs in rural Rajasthan, India, in partnership with a digital health non-profit organization that supports ASHA workers. 
Our goal was to understand how ASHAs respond when beneficiaries raise concerns during family-planning discussions, how these responses relate to ASHAs' own accounts of their communication practices, and what implications these findings hold for AI-supported training.
Each study session combined three components: a simulated phone call with an actor playing a beneficiary raising pre-defined family planning concerns, a semi-structured interview about communication experience and training history, and a brief interaction with an LLM-powered roleplay training chatbot used as a design probe.
This pairing of observed behavior (the call) with self-report (the interview) was deliberate. The scripted calls captured responses to standardized concerns, while the interviews described practices unfolding across repeated visits and established relationships.

On analyzing the call data, we found that ASHAs' responses fell into a small number of recurring categories. Most often, concerns were left unaddressed: participants returned to health-risk information, contradicted the concern without exploring it, promised assistance without specifying a plan, or listed medical solutions (like contraceptive methods) without explaining them. A smaller set of responses instead took up the concern, sought permission before involving family members, or left the decision with the beneficiary. We interpret these moves through Motivational Interviewing (MI), which provides established concepts for understanding both patterns \cite{miller2012motivational}. The chatbot probe surfaced a deployment risk: several participants proposed using the system outside training, by playing its audio to family members or querying it for clinical information.
Each data source contributes differently to the design considerations that follow. The calls reveal the sequential conversational patterns that a feedback system would need to detect. The interviews and design probe reactions supply the constraints such a system would operate under, including workload, connectivity, and the risk of the tool being repurposed as a persuasive voice or informational resource.

We make three contributions.

\begin{enumerate}
    \item Characterize how ASHAs respond to standardized social, material, and health-related concerns in two scripted family-planning scenarios, and compare these responses with participants' accounts of their communication practices.

    \item Identify recurring conversational moves in these responses and interpret them through Motivational Interviewing, treating informed refusal and postponement as legitimate outcomes.

    \item Derive design considerations for LLM-based roleplay training, grounded in observed ASHA communication practices and participants' reactions to a chatbot design probe.
\end{enumerate}

\begin{figure}[h]
    \centering
    \includegraphics[width=0.7\linewidth]{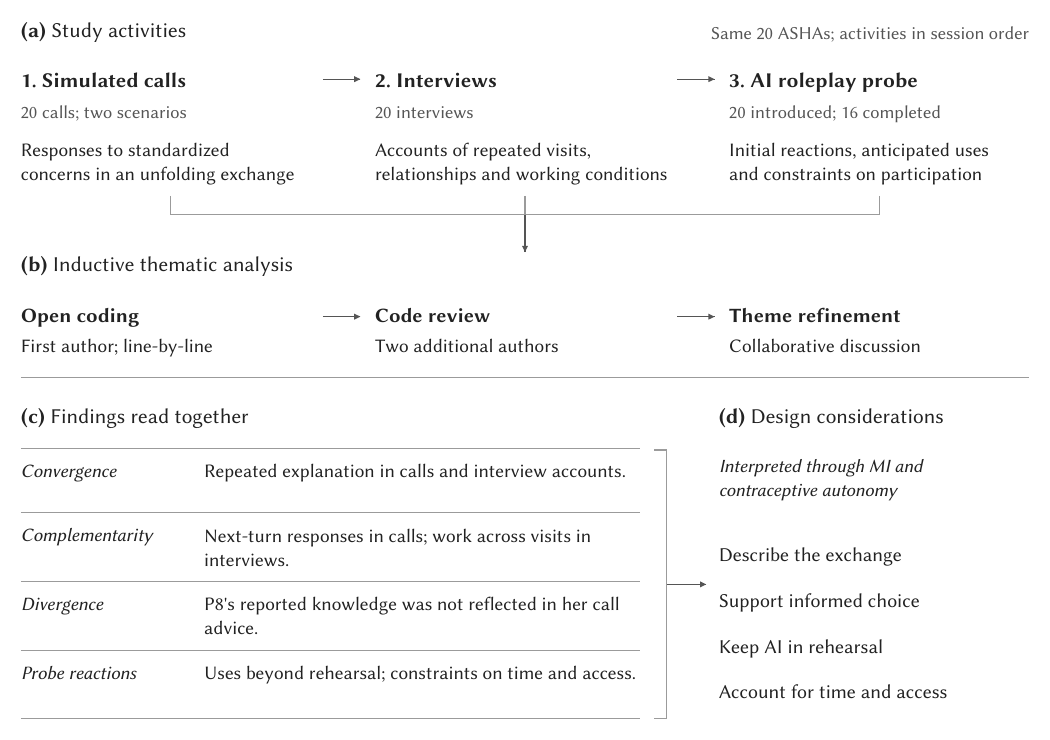}
    \caption{Three activities inform complementary accounts of communication practice}
      \Description{A four-part flow diagram summarising the study design and analysis.
  Panel (a), Study activities, shows three boxes in session order, all with the same
  20 ASHAs: simulated calls (20 calls, two scenarios), capturing responses to
  standardized concerns in an unfolding exchange; semi-structured interviews (20
  interviews), capturing accounts of repeated visits, relationships and working
  conditions; and an AI roleplay probe (introduced to 20 participants, completed by
  16), capturing initial reactions, anticipated uses and constraints on
  participation. Panel (b), Inductive thematic analysis, shows three sequential
  stages: open coding line-by-line by the first author, code review by two additional
  authors, and theme refinement through collaborative discussion. Panel (c), Findings
  read together, lists four relationships between the data sources: convergence,
  repeated explanation appears in both calls and interview accounts; complementarity,
  calls show next-turn responses while interviews show work across visits;
  divergence, P8's reported knowledge was not reflected in her call advice; and probe
  reactions, uses beyond rehearsal and constraints on time and access. Panel (d),
  Design considerations, lists the outputs: findings interpreted through Motivational
  Interviewing and contraceptive autonomy, describe the exchange, support informed
  choice, keep AI in rehearsal, and account for time and access.}
    \label{fig:01}
\end{figure}

Figure~\ref{fig:01} summarizes the study activities, analytic process, and contributions of the three data sources. The synthesis draws on convergence and divergence between calls and interviews, alongside reactions to the chatbot probe, to inform design considerations for AI-supported roleplay training.

\section{Related Work}

\subsection{ASHAs and the Invisible Work of Community Health}
Many digital tools have been built to support CHWs: monitoring and data collection systems for pregnant women and new mothers, supply tracking tools, performance feedback applications, and knowledge-building tools that provide health information via video or messaging platforms \cite{schoeman2019digital,long2018digital}. \citet{ming2022invisible} examined what these toolbuilders may miss: the invisible work ASHAs do alongside their more formal tasks, including relationship-building, boundary-negotiating, and emotional labor involved in working with resistant beneficiaries and unsupportive families. Their study found that much of this work (e.g., efforts to establish trust with a skeptical household, navigate caste and gender dynamics during a visit, persist with a family that has declined care) goes unrecognized and uncompensated. Technology designed for CHWs has tended to focus on data collection and protocol compliance, leaving the relational and persuasive core of the work largely unaddressed.

Human-computer interaction literature provides conceptual vocabulary for understanding why this relational work is so persistently overlooked. \citet{star1999layers} argued that no work is inherently visible or invisible; what counts as work is negotiated through indicators, representations, and institutional definitions. Work becomes invisible when its indicators are absent from the systems that measure and reward performance. In the ASHA context, coverage rates, referral counts, and form completion are all legible to the health system; the communicative labor of persuading a resistant mother-in-law or adapting an argument after it has failed is not. \citet{verdezoto2021invisible} applied this invisible work framework directly to multiple categories of Indian frontline health workers in Karnataka, documenting how their maintenance work, such as anticipating needs, reconciling infrastructure gaps, and supporting care pathways, goes unrecognized by the digital systems designed to support them. They argue for moving beyond the focus on training and performance to support frontline health workers as ``system-builders'' who make healthcare infrastructures work. Our study extends this line of inquiry to a specific form of invisible labor that has not been empirically examined at the level of observed interaction: the communicative work ASHAs do, or fail to do, when a beneficiary or her family resists a health recommendation.

\subsection{AI Tools for Community Health Workers}
The integration of AI into CHW workflows is accelerating. Diagnostic tools, chatbots for health information retrieval, and scheduling systems~\cite{ramjee2025ashabot, smartkc} are all in various stages of development and deployment across the Global South. \citet{okolo2021cannot} conducted an early, pre-LLM study examining how ASHAs actually perceive and understand AI. They found ASHAs had low AI literacy but were relatively receptive to AI tools, yet often attributed infallibility to AI outputs in ways that raised concerns about overreliance. A subsequent study examined how ASHAs interacted with explainable AI interfaces designed to accompany diagnostic predictions, finding that even simple explainable AI visualizations produced widespread misinterpretations \cite{okolo2024if}. More recently, \citet{wadhwa2025designing} conducted a vignette study with ASHAs receiving informational content via LLM-based health chatbots using different messaging frames. They found that while narrative framings were preferred over purely informational framings, narrative framings led to uncritical overreliance in high-ambiguity scenarios.

These studies point to a recurring concern: CHWs are increasingly expected to work with AI tools whose outputs they do not fully understand and whose failure modes they are not equipped to identify. These findings illustrate the importance of preserving workers' judgment and agency, a design priority emphasized by \citet{ismail2021ai} in their thematic discourse analysis of AI for frontline health. This matters for our work in two ways. First, any communication training chatbot will be used by ASHAs whose baseline AI familiarity may be low and whose relationship with digital tools could be shaped, in part, by fatigue and digital reporting burden. Second, the finding that ASHAs may defer to AI outputs, even incorrect ones, has direct relevance to how a training chatbot should be framed. A tool that presents scripted responses as definitive answers, rather than as material to consider and reason about, could produce the same overreliance reported earlier.

\subsection{Existing Work on ASHA Communication}

ASHA training materials center on clinical content (e.g., health protocols, referral pathways, contraceptive methods) with comparatively less attention to how ASHAs should handle resistance, navigate family power dynamics, or adapt their approach when standard methods fail \cite{nhsrc_asha_induction_2021,nrhm2013asha}. Similarly, \citet{yadav2021illustrating} identified limited guidance on breastfeeding counseling in ASHA reference materials and highlighted the need for case-based discussion with experts. Research has found that such underdeveloped counseling and communication skills can undermine ASHAs' self-confidence, leading their home visits to be less effective \cite{shrivastava2016measuring}. Efforts to strengthen ASHA communication training have included mobile videos paired with coaching on asking questions and checking understanding during maternal-health discussions \cite{ramachandran2010mobile}.

Beyond gaps in training support, \citet{goldwater2025community} identify a mismatch between how ASHAs explain beneficiaries' resistance and how they propose responding to it. In a recent vignette study with 146 ASHAs in Bihar, India, they found that when asked why a beneficiary might reject advice, ASHAs frequently cited social dynamics such as family pressure, household norms, and community beliefs. When asked further what they would do to counsel that same beneficiary, they often defaulted to health-benefit arguments, with social dynamic arguments missing from their responses. The authors interpret these findings through the ``information deficit model'', defined as the implicit assumption that resistance is a knowledge problem and that providing better information is therefore the solution.

The information deficit model's persistence matters because it is not wrong (health information does have persuasive value in some contexts) but it is incomplete. Research in communication and behavioral science has documented that information provision alone rarely shifts decisions in the face of social pressure, identity commitments, or household power dynamics \cite{wang2025dismantling,kelly2016changing,seethaler2019science}. 

The model also maps onto a broader pattern that HCI scholarship has long identified: the tendency to design systems around information transfer rather than the coordination, negotiation, and relational work that makes cooperative action possible \cite{schmidt1992taking}. \citet{schmidt1992taking} distinguish between two kinds of effort in cooperative work: doing the task, and doing the work that makes the task possible when the people involved have different goals and constraints. They call the second \textit{articulation work}. ASHA training primarily covers the first, specifically the production work of health service delivery (what information to convey, which protocols to follow). However, it says much less about the second, such as aligning a beneficiary's preferences with her family's constraints, negotiating between institutional health goals and household realities, or adapting one's approach when a standard argument fails.

 Suchman's distinction between plans and situated actions provides a complementary lens \cite{suchman1987plans}. The information deficit model treats communication as a plan: deliver health information in the prescribed sequence, and compliance follows. When resistance arises, the prescribed response is to re-execute the plan more carefully or persistently. But effective communication with resistant beneficiaries is fundamentally situated. It requires reading the specific social dynamics of a household in the moment and improvising a response to what the beneficiary actually says. Suchman argued that plans serve as resources for action rather than determinants of it, and that a system designed on the planning model will be insensitive to the particular circumstances of the encounter. This theoretical frame motivates our study's behavioral methodology.

We build on \citet{goldwater2025community}'s identification of the information deficit model by observing what it produces in a more naturalistic, live setting with actual social pressure, in conversations with simulated beneficiaries. Their data is vignette-based (i.e., ASHAs reasoning about hypothetical scenarios). Our work provides a behavioral complement. The simulated phone calls in our study let us observe what ASHAs might say when a caller pushes back, or raises a concern the ASHA's standard script does not cover. 

\subsection{Motivational Interviewing and Roleplay-Based Communication Training}

Motivational Interviewing (MI) offers an established approach to conversations about health-related change. It emphasizes collaboration, exploration of ambivalence, and eliciting a person's own motivations through practices such as open questions, reflective listening, and affirmations~\cite{miller2012motivational}. MI also identifies the \emph{righting reflex}: a practitioner's impulse to correct, persuade, or supply solutions when a person hesitates. Its relevance to our study lies in these observable conversational practices, which provide vocabulary for examining how workers respond to concerns and offer information or advice. Although MI is oriented toward change, it emphasizes respect for the person's autonomy and negotiation of the conversation's goals~\cite{miller2012motivational}.

MI training has previously been adapted to frontline workers. For example, in South Africa, \citet{mokhele2024improving} evaluated a training and support program for lay HIV counsellors that combined initial instruction, roleplay, review of recorded encounters, feedback, and quarterly refresher sessions over twelve months. They reported improvements in technical MI skills among program completers. This work motivates investigating how repeated practice and feedback might support communication skill development among frontline workers, while leaving adaptation to ASHAs and family-planning counseling an empirical question.

Roleplay provides one setting for practicing these conversational skills. Communication training using standardized patients, in which a trained actor portrays a patient with specified concerns, has been associated with improvements in communication and self-confidence~\cite{xiao2025use,ross2024impact}. However, its delivery is constrained by costs, actor availability, and the resources required to maintain training quality~\cite{NewlinCanzone2013,Hart2016,Brenner2009,Nagpal2021}. LLMs now open a pathway toward scaling consistent roleplay training at low cost. For instance, recent HCI work has explored how roleplay with LLMs can support clinician and counselor education in the United States \cite{sehgal2025pal,baseman2025poker,haut2025ai,steenstra2025scaffolding}. However, the extension of this method to community health settings in the Global South is less developed. Voice interaction, low data bandwidth, digital tool fatigue, low tech literacy, and the specific social dynamics of ASHAs all impose requirements that systems built for other populations, often in the Global North, may not address \cite{sehgal2025exploring}. Importantly, before such a tool can be responsibly built, the communication practice it is meant to train must be empirically characterized.

\subsection{Family Planning, Coercion, and Contraceptive Autonomy in India}

India's history of coercive population control and concerns about targets and incentives make it especially important to examine what counts as a successful counseling encounter \cite{gwatkin1979political,williams2014storming}. Female sterilization is the largest single contraceptive method in India and ASHAs receive per-case payments for counseling, motivating, and following up with sterilization clients \cite{iips2026nfhs6,nhm2025ashaincentives}.
 

 
Against this background, family planning scholarship distinguishes the quality of counseling from its contraceptive outcome. \citet{bruce1990fundamental}'s quality-of-care framework treats choice of method, information given to clients, and interpersonal relations as dimensions of quality to be evaluated in their own right rather than inferred from contraceptive uptake. More recent work defines contraceptive autonomy in terms of whether a person can make and act on an informed and voluntary contraceptive decision, including a decision not to use contraception \cite{senderowicz2020contraceptive}. Contraceptive coercion can also take subtler forms than overt force, including directive counseling, scare tactics, and constrained method choice \cite{senderowicz2019obligated}. Research from Gujarat similarly documents provider guidance toward particular methods and gaps in counseling about contraceptive options and side effects \cite{holt2021gujarat}.

Taken together, a frontline worker who becomes more effective at persuasion is therefore not, by that fact alone, counseling better. The relevant question is whether the conversation improved a woman's understanding and supported her ability to make her own decision, including a decision to delay or decline contraception. Section 5 returns to what this implies for the design of training tools.

\section{Methodology}

\subsection{Study Protocol}

We conducted a qualitative study with 20 ASHAs from 12 villages in 
a district in Southern Rajasthan, India. Participants were recruited through a digital health non-profit organization that works closely with ASHAs in the region. Participation was voluntary. All sessions took place in person. Most sessions (18) were conducted at Anganwadis (local rural health centers), which allowed us to minimize disruption to their workdays and conduct the study in a familiar setting. Two sessions were conducted in participants’ homes. Participants had 7-21 years of experience as ASHAs (mean= 13.4). Appendix Table 2 provides a participant overview.

Three researchers were present for the study: a female author fluent in Hindi led the study activities, a second female author fluent in Hindi took notes, and a female author from the non-profit partner facilitated introductions and helped establish rapport. All sessions were conducted in Hindi, and took 45-60 minutes.

Each session followed three steps in sequence: (1) an unannounced simulated phone call 
(2) a semi-structured interview, and (3) an interaction with a chatbot design probe. We describe each component below. The complete protocol, including the use of a  simulated call and audio recording before disclosure, was reviewed and approved by the Institutional Review Board.

Participants were not compensated for study participation, following guidance from our non-profit partner. However, they received reimbursement for mobile data costs from the non-profit. This support was independent of study participation and was not treated as research compensation. We discuss the limitations of delayed disclosure, partner-mediated recruitment, and non-compensation in the Limitations section.

\subsubsection{Simulated Call}
Each ASHA participated in a scripted phone call in which a member of the non-profit played the role of a beneficiary seeking advice. Participants were randomly assigned to one of two scenarios (10 participants per scenario). The first scenario focused on birth spacing. The caller had a nine-month-old daughter and was seeking contraceptive guidance while facing resistance from family members. She was also wary of the Antara injection, a contraceptive introduced by the Government of India in 2017 that provides three months of protection per dose. Although effective and widely promoted by ASHAs, women commonly express concerns about side effects such as menstrual irregularities, which can affect acceptance and continuation \cite{ray2024experiences,agrawal2022insights}. The second scenario focused on sterilization and preference for a son. The caller had four daughters and was facing pressure from her family to continue childbearing in the hope of having a son. It reflects well-documented patterns in which reproductive decisions are strongly influenced by husbands and in-laws in contexts where son preference remains culturally entrenched \cite{oliveira2014dominance}. Both scenarios were informed by prior literature, and were decided and refined in consultation with ASHA supervisors and the non-profit team. To ensure comparability across participants, the beneficiary actor followed a structured script with predefined resistance points and objections to raise if the ASHA did not address them. The same actor conducted all calls.

The call was conducted unannounced. At the start of the session, a member of the non-profit handed her phone to the ASHA and indicated that a beneficiary was seeking guidance. This design involved deception for methodological reasons. 
Informing participants in advance that the call was part of a study could alter how they responded to concerns in the moment. 
Unannounced standardized or ``mystery'' clients are an established method for assessing counseling and service quality in health-systems research, including in India \cite{das2016impact, allotey2025using}.
The scripted call was intended as a standardized elicitation instrument, and not a proxy for a home visit. Holding the caller's concerns constant across participants sacrifices some naturalism and reduces contextual variation in exchange for comparability, a tradeoff standardized-patient methods make deliberately ~\citep{cox2023toolkit}.

Because the call was unannounced, informed consent for study participation and use of the recording was obtained immediately after the call. Participants could decline participation, in which case the recording was deleted. The call was placed first in the study sequence so that communication behavior could be observed before the interview prompted participants to reflect on their practices.

\subsubsection{Semi-Structured Interview}
Following the call, the other two members of the research team entered the Anganwadi and joined the session. Informed consent to participate was obtained at this point. Each ASHA completed a semi-structured interview covering: their general experience communicating with beneficiaries, the topics they found most difficult to discuss, the strategies they used when beneficiaries or family members resisted their recommendations, their experience with existing digital tools including ASHA Saheli (an LLM-based WhatsApp chatbot that answers their questions~\cite{ramjee2025ashabot}), and their training history. The interview guide was developed iteratively and piloted with two ASHAs prior to fieldwork.

\subsubsection{Chatbot Design Probe}
In the final study activity, ASHAs were introduced to a communication-training chatbot design probe. The chatbot resembled a WhatsApp interface and was deployed as a web application. It offered two scenarios mirroring the simulated call scenarios: Antara injection hesitancy during birth-spacing counseling (Scenario A) and sterilization in the context of son preference (Scenario B). Each ASHA selected one scenario and completed a single interaction with the chatbot. Following the interaction, ASHAs were asked about the realism of the simulated beneficiary, the usability of the system, and whether they could imagine using a similar tool in their work.

\begin{figure}[t]
\centering
\begin{subfigure}{0.35\linewidth}
    \centering
    \includegraphics[width=\linewidth]{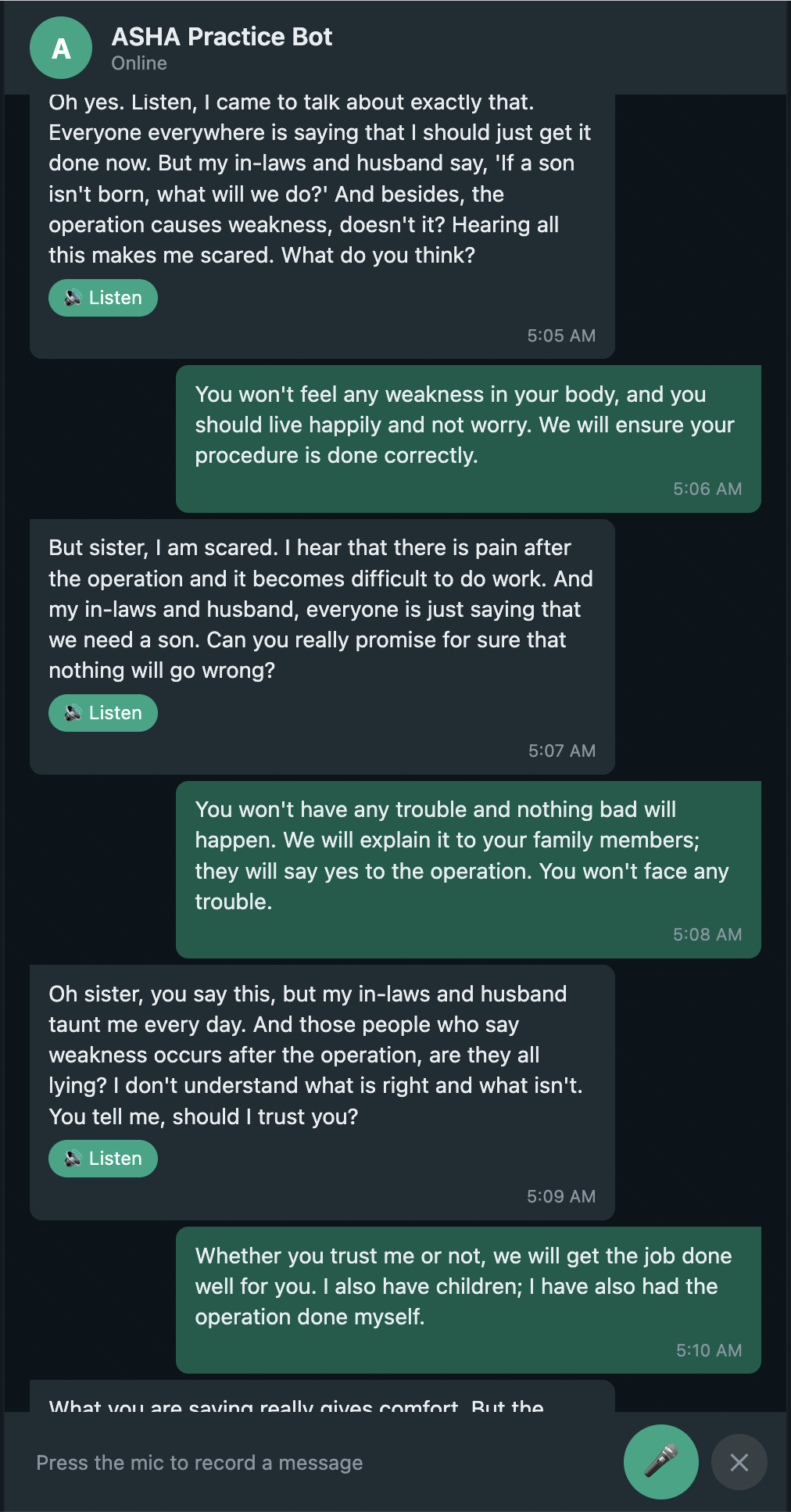}
\end{subfigure}
\caption{Chatbot design probe during P5's interaction, translated into English.}
\Description{Chatbot design probe during P5's interaction, translated into English.}
\Description{A screenshot of the chatbot design probe during P5's interaction, shown
in a dark-themed messaging interface resembling WhatsApp and headed "ASHA Practice
Bot, Online". The conversation, originally in Hindi, is translated into English.
Messages from the simulated beneficiary appear on the left, each with a "Listen"
button for audio playback; P5's replies appear on the right. The beneficiary says
everyone is telling her to have the sterilization done, but her in-laws and husband
ask what they will do if a son is not born, and that she has heard the operation
causes weakness, which frightens her. P5 replies that she will not feel weakness and
should live happily, and that the procedure will be done correctly. The beneficiary
says she is scared, has heard the operation is painful and makes work difficult, and
asks whether P5 can promise that nothing will go wrong. P5 replies that she will have
no trouble and that the family will be persuaded to agree. The beneficiary says her
in-laws and husband taunt her daily and asks whether the people describing weakness
are lying and whether she should trust P5. P5 replies that whether or not she is
trusted, the job will be done well, and that she has children herself and has had the
operation done. A microphone button for recording a reply is visible at the bottom of
the screen.}
\label{fig:chatbot_prototype}
\end{figure}

The probe was implemented as a browser-based voice chatbot using OpenAI’s GPT-4o. Participant speech was transcribed using OpenAI Whisper (\texttt{whisper-1}, configured for Hindi), appended to the conversation history, and passed to GPT-4o to generate the beneficiary's response. Responses were displayed as text and synthesized into speech using Google Cloud Text-to-Speech (\texttt{gemini-2.5-flash-lite-preview-tts}, Aoede voice). Figure~\ref{fig:chatbot_prototype} shows the chatbot interface during P5’s interaction, with the conversation translated into English.

\subsection{Data Collection and Analysis}
The study generated three data sources per participant: a simulated call recording, a semi-structured interview recording, and a chatbot interaction log. All audio recordings were translated from Hindi into English and transcribed using Gemini 3 Pro Preview. We evaluated multiple transcription and translation pipelines, including models from OpenAI and Sarvam AI, and selected Gemini based on output quality. To assess transcription reliability, one interviewer reviewed two randomly selected interviews and two call transcripts, and identified no substantive issues. In addition, all direct quotes reported in this paper were verified against the original Hindi audio.

Transcripts and logs were analyzed using inductive thematic analysis \cite{braun2006using}. The first author reviewed all transcripts and logs, and conducted line-by-line open coding. The resulting codes were then reviewed by two additional authors, and the three of them collaboratively identified emerging patterns and themes. 
Through iterative discussion and refinement, these codes were consolidated into higher-level themes that informed the findings reported below.

\subsection{Ethics and Positionality}
The research team comprised seven members. All are of Indian origin; all but one were born and raised in India. Two team members work for a digital health non-profit that builds tools for ASHA workers. While shared cultural and linguistic backgrounds may have aided rapport and interpretation, we note significant structural differences between the research team and participants. ASHAs are low-income rural workers operating within a hierarchical public-health system, whereas the research team occupied positions of institutional authority. Several participants showed signs of discomfort, appearing to view the study as a form of performance evaluation on behalf of a government body. This perception likely shaped what they felt comfortable disclosing. The simulated calls and interviews were differently, not unequally, shaped by this dynamic. Interviews were shaped by self-presentation and perceived evaluation, whereas calls were shaped by surprise, the constraints of a brief telephone interaction, and the same perceived evaluation (Section~4.2). We therefore treat neither data source as ground truth.
Instead, we attend analytically to points of convergence and divergence between observed behavior and participants' accounts of their own practice.

\section{Findings}

We organize the findings into six sections. Section 4.1 describes recurring patterns in which ASHAs returned to previously shared health information without directly engaging the concern raised by the simulated caller. Section 4.2 examines differences between participants’ interview accounts and their responses during the scripted calls. Section 4.3 characterizes the social, material, and health-related concerns elicited by the two scenarios. Section 4.4 identifies concern-addressing conversational moves and interprets them through the lens of Motivational Interviewing. Section 4.5 reports how participants recalled and described their prior training. Section 4.6 presents participants’ reactions to the chatbot design probe and the constraints they associated with existing digital tools.

\subsection{Repetition and Concern-Evading Responses in the Scripted Calls}

Across the simulated calls and interviews, repetition emerged as the most common counseling strategy.
When beneficiaries did not accept a recommendation, participants often described explaining the same information multiple times.
In the scripted calls, repetition sometimes took a more specific form: when callers introduced a social, material, or health-related concern that was not resolved by the ASHA's initial answer, ASHAs often returned to an earlier health-risk or gender equity argument rather than exploring the concern directly. We refer to these as concern-evading responses.

In interviews, ASHAs described repetition as a routine response when community members continued to decline a recommendation. P1 recounted visiting the same household repeatedly:

\begin{quote}
“Four-five times... just to explain to them. Then we also tell them that medicines [contraceptive pills] cause harm, the uterus can get damaged by medicines, you get the operation done. Still it has no effect on them. They just don’t get it done...We are tired of explaining, they just don’t understand.”
\end{quote}

P3 characterized repetition as the recommended strategy based on her formal training: “If they don’t understand once, we will go a second time, third time... we will try to explain so they understand... We have to keep trying like this.”

\subsubsection{Pivoting to Risk Language}
A common concern-evading trajectory involved shifting from a social or material concern to a general health-risk claim. P1’s caller repeatedly framed the family’s desire for a son as an old-age-security concern: 

\begin{quote}
\textbf{Caller:} Yes, there is risk, Madam, but there should be at least one boy in the house among all the brothers-in-law.

\textbf{P1:} It should be, but now that you have four daughters, the girls are the boys now. Understand it like that.

\textbf{Caller:} But Madam, the girls will marry and leave. Then who will support us in old age? Everyone needs one boy. I also hope that I might have a boy.

\textbf{P1:} But this is a risky matter. Because there are four girls---who knows what will happen next?

\textbf{Caller:} Yes Madam, I understand, but my husband and mother-in-law also want me to have a boy.

\textbf{P1:} Explain to them that the girls themselves are the boys now. After four girls, going forward is a risky matter.
\end{quote}

Notably, the caller did not dispute that another pregnancy could be risky. Instead, she repeatedly returned to old-age support and pressure from her husband and mother-in-law. P1 alternated between health-risk and gender-equity arguments rather than taking up those constraints. 

This pattern is consistent with the distinction identified by \citet{goldwater2025community} between recognizing a social constraint and subsequently responding with health information. Our data extend that observation by showing how the mismatch can persist across a multi-turn exchange. They do not, however, establish why participants returned to these arguments. We propose possible explanations in Section 4.5.

\subsubsection{Categorically Denying a Concern}
A second pattern was categorical denial: rejecting a concern without investigating or substantiating the rejection. When P1’s caller said that her family wanted a son and would not agree, P1 replied, ``They will agree; you talk to them.'' When P8’s caller described reports of menstrual irregularities after Antara, P8 responded, ``No, no, there is no problem. You can get Antara as well, there is no problem.'' Neither participant asked what the caller had heard or experienced, acknowledged why the concern might be credible, or explained the basis for reassurance. The denial replaced inquiry with an unsupported counter-assertion.

\subsubsection{Offering Reassurance Without a Plan}

A third pattern appeared when callers explicitly requested help constructing an argument to use with resistant family members, rather than additional information. The call with P5 provides a clear illustration. The caller had already tried to convince her family and was seeking guidance on what to say next:

\begin{quote}
\textbf{Caller:} Madam, that is what I want from you. It is your day and night job. Explain to me how I should convince my husband and mother-in-law.

\textbf{P5:} We will explain.

\textbf{Caller:} You help me, madam, how to explain to them.

\textbf{P5:} No, we will make them listen. We will explain to them. Yes, yes, we will definitely explain.

\textbf{Caller:} \textit{*Pauses with a brief moment of silence*}

\textbf{P5:} Okay. Is there any other problem?
\end{quote}

Here, the caller precisely stated what she needed: not clinical information, but an argument she could use with a family that remained unconvinced. 

P5 responded with repeated assurances that help would be provided, but without specifying what that help would involve, followed by a move to end the call. 

\subsubsection{Naming Methods with Limited Explanation}
All 10 ASHAs assigned to the birth-spacing scenario named at least one contraceptive method; most named two or three. However, methods were more often listed than explained. Dosage instructions were frequently incomplete or inconsistent. Individualized guidance was rare. P7 admitted during the call that she had forgotten details about Chhaya (a weekly contraceptive pill) after having already recommended it. 

P17's call showed the same gap from the other direction: rather than forgetting details about a method she had recommended, she declined to discuss one she could not explain.

\begin{quote}
\textbf{P17:} Besides Nirodh [male condom brand], there is nothing better.

\textbf{Caller:} Someone also told me about some pill\ldots

\textbf{P17:} Regarding the pill, ma'am, I do not know much about its use. Although they give it, I do not have much information about it.
\end{quote}

P17 then returned to recommending condoms. The exchange illustrates the difference between naming a method and counseling about it. Once the caller requested method-specific guidance, P17 could not explain how the pill was used, what the caller might expect, or how it compared with the alternatives she had named.

P6 provided a contrasting case: she described several contraceptive options, connected birth spacing to recovery, offered continued contact, and offered to speak with the caller's mother-in-law if additional concerns arose. This case shows that method-rich counseling involves comparison, explanation, and follow-through rather than a long list of medication names.

Taken together, these patterns constitute four recurring forms of concern-evading responses: shifting from a stated concern to risk language, denying the concern without explanation, promising help without specifying a plan, and listing contraceptive methods with limited explanation or individualized guidance. These patterns operated at different interactional scales. Categorical denial could occur in a single reply, whereas risk pivots and reassurance without a plan often became visible across several turns. We return in the Discussion to the implications of these patterns for AI-supported training and feedback.

\subsection{What Interviews and Simulated Calls Made Visible}
Participants often began the interview with broad and favorable descriptions of their communication practices, yet later in the same conversation recounted detailed examples of difficulty in communication. We interpret this pattern as a difference between general self-description and incident-based reflection rather than simply a denial or contradiction. It may reflect social desirability, perceived evaluation, different interpretations of what constitutes a communication difficulty, or the greater specificity elicited by follow-up questions. A further possibility, developed in Section 4.5, is that participants' training does not frame these episodes as communication difficulties at all. If persistence through repeated explanation is understood as the correct practice, a family that has not yet agreed may be viewed as an ongoing case rather than a communication problem.

For instance, when asked whether she faced communication difficulties, P17 responded: ``No, nothing like that happens. In fact, villagers listen to me. Even when the ANM [Auxiliary Nurse Midwife] goes, they don’t have as much connection with her.'' Later in the interview, however, she described repeated unsuccessful attempts to counsel a family that continued trying for another pregnancy in the hope of having a son. Similarly, P10 initially said that “such a time hasn’t come where people didn’t understand... in all these years I’ve worked,” but later described several strategies she used when families continued to decline her recommendation.

The scripted call data revealed a different (but not inherently more truthful) view of ASHA communication practice. Participants readily named contraceptive methods, physiological rationales, and health risks, but less often asked follow-up questions about the social and material constraints introduced by the caller. By contrast, the interviews produced descriptions of communication strategies extending across repeated visits, established relationships, and interactions with family members, which short scripted calls cannot fully reproduce.

P9 was the only participant who explicitly described planning a different approach when an earlier attempt had not succeeded. She explained: “That remains in the mind... ‘this time I said this and she isn’t agreeing’... so we go thinking of something else.” Her account suggests reflective adaptation across encounters. However, this strategy was only described in P9’s interview and did not appear in her simulated call.

Notably, one participant showed a direct discrepancy between the two activities. During the call, the caller asked whether becoming pregnant again, with a nine-month-old at home, could be harmful, P8 replied: ``No, no, there is no harm.'' The caller asked twice; P8 denied harm both times. In her interview, however, P8 correctly described the importance of birth spacing and the risks associated with becoming pregnant too soon after a previous birth. In this case, knowledge that P8 articulated during reflection was not reflected in the advice she gave during the call. The call and interview data are therefore in direct contradiction: accurate knowledge did not transfer to accurate advice under the pressure of live interaction.

Rather than treating either data source as a ground-truth account of routine practice, we view them as capturing different facets of ASHA communication under different conditions.

Interviews may have been shaped by social desirability, researcher expectations, and accumulated experience. They represented counseling as longitudinal and distributed. Participants described returning over several days, building trust through assistance beyond family planning, consulting ANMs, and speaking separately with women and family members. The calls instead required an immediate response to an unfamiliar caller’s standardized objection. The calls may have been shaped by surprise, unfamiliarity with the caller, the constraints of a brief telephone interaction, and variation in the actor’s enactment. The difference is therefore partly one of temporal and relational scale. Interviews made coordination across encounters visible, while calls made the next response in a difficult exchange visible.

\subsection{Social, Material, and Health-Related Concerns in the Two Call Scenarios}
The two scenarios elicited concerns extending beyond factual knowledge, including household influence over reproductive decisions, preference for a son, recovery-related caregiving and domestic work, and community circulated accounts of contraceptive side effects. These concerns combine social, material, and health-related considerations in different ways. Notably, while the scenarios represent two specific configurations of family-planning decision making, they illustrate the broader contexts in which women accept, postpone, or decline family-planning services.

\subsubsection{Sterilization, Son Preference, and Household Power}
In interviews, participants most often described sterilization counseling as difficult when husbands or in-laws opposed the procedure in pursuit of a son. P16 summarized the leverage of household authority: “If the husband agrees, then the whole family becomes easy to work with.”

P16's account illustrates how reproductive decisions are shaped by household power relations. Agreement from a husband or mother-in-law is not equivalent to the woman’s own free and informed choice.

Participants described several ways of responding to son preference. The most common was a gender-equity argument, such as “boys and girls are equal now” (P13). Some highlighted government benefits available to families with daughters: “if there are girls, you will get even more facilities from the government” (P2). Others emphasized that another pregnancy could not guarantee a son: “if a boy comes after the girls it’s great, but if a boy doesn’t come and another girl comes, then what will you do?” (P13).
These responses treated son preference primarily as a belief to counter. They less often engaged the specific concerns attached to it in the calls, including old-age support and the caller's limited influence over household decisions.

P12 offered a different response. When the caller raised concerns about support in old age, P12 replied: “Actually, it is the girls who serve and care more.” This statement engaged the old-age care concern on its own terms. At the same time, it relied on another gendered expectation, that daughters should provide family care. 

These responses addressed son preference as an argument to rebut, while the scenario presented it as a constraint on the caller’s ability to act. The unresolved question was not only whether daughters and sons were equal, but what support the caller wanted when her husband and in-laws could constrain her reproductive decision.

\subsubsection{Household Work and Recovery}
Callers also framed sterilization recovery as a labor-allocation problem: who would cook, care for children, and perform physically demanding work while the woman recovered? This worry was distinct from disagreement about the clinical value of sterilization because it concerned whether the household could absorb the practical consequences of the procedure.

P6 described an incident in which she had told a community member: “If you get the operation done, I will make \textit{roti} [whole wheat flatbread] for you.” This offer directly addressed the practical barrier being raised, but it also illustrates how ASHAs may attempt to absorb additional domestic and care labor to make health services feasible.

In the scripted calls, however, most participants responded with general reassurances rather than practical planning. P5 stated that “there is no trouble from the operation, everyone gets it done.” P16 challenged the premise of the concern: “It’s not like after the operation we can’t do anything. These are all mental assumptions.” 

P12's call illustrates how this material concern unfolded across several turns:

\begin{quote}
\textbf{Caller:} If I get the operation done, I will have to stay at home and rest. Then who will look after the children in my absence?

\textbf{P12:} There is nothing like that in the operation.

\textbf{Caller:} So one does not have to rest?

\textbf{P12:} One has to rest, but it can be done.

\textbf{Caller:} For how many months does a woman have to rest?

\textbf{P12:} Up to one month is easily fine. For many days, one should not do heavy lifting or such work.

\textbf{Caller:} One month? Then who will work at my home? If so, I cannot get it done right now.

\textbf{P12:} Your eldest daughter will do a little, right? The in-laws are there; they will also do it.
\end{quote}

The caller’s follow-up turned an abstract question about recovery into a concrete labor constraint. P12 responded by assuming that the caller’s eldest daughter and in-laws could absorb the work. The response therefore proposed a solution without establishing whether that household support was actually available.

P12 also estimated a one-month recovery period, but other participants gave substantially different estimates. Reported recovery periods ranged from 8--10 days to two months. This variation points to inconsistency in clinical guidance rather than a communication skill gap. No participant marked the estimate as uncertain, explained what recovery time depended on, or suggested checking individualized guidance with a clinician.

The issue was therefore both clinical and material. Recovery time determined whether the caller could arrange childcare and household work. Unqualified and inconsistent estimates left her without a reliable basis for planning, even when the advice was intended to reassure her.

\subsubsection{Antara and Community-Circulated Concerns}
In the birth-spacing scenario, the caller raised concerns about Antara, including reports that menstruation might stop or become unusually heavy. Participants responded in two broad ways. Some acknowledged that menstrual changes could occur, explained that they were associated with the method, and discussed continuation strategies or alternatives. Others categorically contradicted the concern. P4, for example, said: ``No, nothing like that will happen,'' despite menstrual irregularities being a well-documented side effect of Antara.

These denials set the ASHA’s assertion against accounts circulating among the caller’s neighbors and family without explaining why one account should be trusted over the other. The interaction therefore left both the side-effect question and the credibility conflict unresolved.

\subsection{When ASHAs Engaged the Concern}
Alongside the concern-evading patterns described in Section~4.1, the calls and interviews included examples of engaging a stated concern, making offers of support conditional on the woman's wishes, or preserving opportunities for further discussion. These examples were concentrated in four participants (P6, P9, P10, P12), although some appeared only in interview accounts. Drawing on the MI concepts introduced in Related Work, we examine how these responses attend to concerns, seek collaboration, and support choice. We distinguish addressing the topic of a concern from reflecting the caller's perspective or supporting her agency: a response can engage the concern while still advancing a counterargument or assuming what the woman wants.
 
Notably, no participant had received MI training, and none displayed MI's full repertoire. We therefore classify individual moves as MI-consistent or MI-inconsistent, but do not classify participants as practicing MI.

\subsubsection{Engaging the Stated Concern}
Where the most frequent responses to the old-age-security concern were contradiction or a return to equity and risk arguments, P12 answered it on its own terms: ``Actually, it is the girls who serve and care more.'' In MI terms this is a reflective move. It takes the family's stated worry seriously and reframes it within their own framework rather than opposing it with an abstract principle. The turn is also instructive about the limits of a single move. It substitutes one gendered expectation (daughters as caregivers) for another, and an MI-consistent continuation would go further such as evoking the caller's own reading of the concern (for instance, asking what the family fears would happen without a son) before supplying any counter-claim.

\subsubsection{Permission, Conditional Offers, and the Woman's Voice}
The family-engagement offers differed not only in specificity but in their stance toward the caller's agency. For example, P6's offer was conditional on the caller's wish: ``\emph{If you want}, make me talk to them, I will explain to them, your mother-in-law, make her talk to me.'' The conditional structure functions as permission-asking, an MI-consistent behavior that keeps the caller in the position of deciding whether, when, and how her family is approached. By contrast, ``we will make them listen, yes yes, we will definitely explain'' (P5) and ``yes I will
explain, okay'' (P19) commit the ASHA to act without eliciting the caller's preference about family involvement at all. P10's interview account extends the permission-keeping stance across encounters: she described convening the woman, her husband, and her mother-in-law together, then maintaining a private telephone channel when the woman could not speak freely in front of her family. This plan treats the
beneficiary's unconstrained voice as something to be protected within the
coordination rather than an obstacle to it. At the other end of the range, P17 described leaving family conversations to the women themselves: ``The women themselves explain to the husband. I don't
explain.'' Our data cannot classify this stance. It may reflect deference to the woman's preferred channel, or disengagement from the hardest part of the coordination.
 
\subsubsection{Closure, Preference, and Legitimate Endpoints}
In 15 of the 20 calls, the interaction ended with the burden returned entirely to the caller (``I will talk more at home''); five closed with a specific next step, an invitation for continued contact, or an offer of support. For example, P6 ended with: ``You can call anytime, any work, anything to ask, anything, you can call anytime. You take my number from them.'' We deliberately do not treat the absence of an agreed next step as an inherent failure: postponement, informed refusal, or a decision not to involve the family
can each be appropriate outcomes, and MI itself defines success by whether the outcome reflects the person's own considered preference, not by uptake~\citep{miller2012motivational}. However, the gap our data show is prior to that question. In most calls, the caller's own preference was never
explicitly elicited, so whatever closure occurred could not be anchored to it.
 
\subsubsection{What These Cases Do and Do Not Establish}
Read together, these moves are articulation work~\citep{schmidt1992taking}: coordinating health information, household constraints, privacy, material realities, and the preferences of several actors whose interests do not automatically cohere. They are also few, unevenly distributed, and in two of four participants visible only in the self-reported interviews, not calls. We therefore argue MI-consistent moves occur in this population, without MI training, alongside a more common repertoire consistent with what MI describes as the righting reflex.

Three clarifications are important. First, we do not claim these four ASHAs possessed a fully articulated model of MI. What we observe is a consistent pattern of moves, not a self-reported framework. Second, MI-consistent moves appeared unevenly even within these participants. P12's reframing move was otherwise uncharacteristic of her call. P10 described her multi-step family engagement plan in interview but did not execute it in the call. Third, years of experience did not predict which ASHAs demonstrated this orientation. P6 had 10 years of experience, and more experienced ASHAs were not systematically more likely to demonstrate MI-consistent moves. This suggests the skill is not a natural product of field experience, and is unevenly distributed.

\subsection{Training as Recalled and Enacted}
Participants generally recalled their training in greater detail when discussing health content than when discussing communication under household or social constraint. They described learning about contraceptive methods, referrals, danger signs, and government programs. Few recalled specific instruction on adapting a response after a stated concern remained unresolved or on supporting a woman whose preferences conflicted with those of her family.

Several participants described repeated explanation as part of the guidance they remembered. P2 stated: “They told us to go once, if they don’t understand once then explain to them a second time... just keep trying like this.” P3 gave a similar account. Asked whether she had received guidance on difficult conversations, P8 said: “No, this wasn’t taught.” Other participants, including P6 and P7, recalled more general guidance to speak with affection or establish rapport.

We situate these recollections alongside relevant guidance in the national
ASHA induction materials~\cite{nrhm2013asha,nhsrc_asha_induction_2021}.
Table~\ref{tab:curriculum_recall} distinguishes selected curriculum content
from the training recollections reported in our interviews. The materials
explicitly address active listening, exploration of contributing factors,
and negotiation, providing context for interpreting what participants
remembered and emphasized.

\begin{table}[t]
    \centering
    \caption{Selected national curriculum guidance and participants'
    training recollections. Examples are illustrative rather than
    exhaustive participant counts. This comparison does not establish
    individual exposure to particular materials or measure retention.
    Page numbers refer to the printed pages of the national induction
    module~\cite{nrhm2013asha}.}
    \label{tab:curriculum_recall}
    \small
    \setlength{\tabcolsep}{4pt}
    \renewcommand{\arraystretch}{1.15}
    \begin{tabular}{@{}
        p{0.16\linewidth}
        p{0.36\linewidth}
        p{0.42\linewidth}
        @{}}
        \toprule
        \textbf{Area} &
        \textbf{National curriculum guidance} &
        \textbf{Recollections reported in this study} \\
        \midrule

        Health information and services &
        The family-planning chapter explains contraceptive methods
        and access to services. &
        Participants recalled contraceptive methods, referrals,
        danger signs, and government programs. \\
        \addlinespace

        Active listening &
        Listening includes checking meaning, paraphrasing, and
        attending to feelings (pp.~32--33). &
        P6 and P7 recalled general guidance about affection and
        rapport; the reported recollections provided limited detail
        about specific listening practices. \\
        \addlinespace

        Exploring concerns &
        Counseling involves examining contributing factors and
        discussing choices with the person (p.~34). &
        P8 said difficult conversations had not been taught.
        Few participants recalled guidance for responding when
        women's preferences conflicted with family expectations. \\
        \addlinespace

        Adapting a response &
        Negotiation includes eliciting concerns, preparing
        alternatives, and avoiding argument (pp.~35--36). &
        P2 and P3 recalled repeated explanation as a recommended
        response when an initial explanation did not succeed. \\

        \bottomrule
    \end{tabular}
\end{table}

This comparison identifies a difference between documented guidance and
what participants emphasized when recalling training. However, we cannot 
establish which materials participants encountered, how local trainers
presented them, or what opportunities participants had to practice these
skills. Local delivery, supervisory practice, retention over time,
workplace incentives, and perceived evaluation during the study may each
have shaped their accounts.

We therefore use the information-deficit model as a sensitizing concept:
a lens that directs attention to particular patterns without establishing
their cause~\cite{bowen2006grounded}. The combination of content-focused
training recollections and repeated information delivery in the scripted
calls is consistent with an information-centered approach to counseling.
Other explanations remain plausible, including reliance on familiar
health messages during an unexpected call, reluctance to probe sensitive
household matters, or attempts to provide an answer perceived as safe
under evaluation.

Our data thus show convergence between participants' emphasis on health
content and repeated explanation in their training recollections and
their use of health-information arguments in the scripted encounters.
The contribution of formal curricula, local training delivery,
supervision, and working conditions to these patterns remains unresolved.

\subsection{Initial Reactions to the Chatbot Probe and Existing Digital Constraints}
All 20 participants were introduced to the chatbot design probe. Network disruptions affected four interactions. Of the 16 participants who completed an interaction, 14 made favorable comments about the realism of the simulated caller or the ease of using the interface. When asked what she liked most, P5 pointed to the simulated beneficiary's account of family pressure to have another child and worries about managing household work. P12 described the exchange as similar to conversations in the field, explaining that
she responded as she would when speaking with women in the village. P6 appreciated the spoken interaction, explaining that small text on a phone was often difficult to read for her. P16 felt the tool would be useful for new ASHAs with less experience. 

Because each participant encountered the system once, with researchers present, we interpret these comments as initial reactions rather than evidence of usability, adoption, realism, learning, or training effectiveness. Novelty and participant response bias may also have shaped the favorable assessments, a bias well-documented in HCI field studies with underserved populations \cite{dell2012yours}.

Still, the overall surface-level reception was encouraging. However, a more careful reading of the chatbot reactions surfaces deployment risks worth naming directly.

\subsubsection{Participant-Envisioned Uses Beyond Roleplay Training}
Two ASHAs proposed using the chatbot audio as a persuasion device, playing the AI beneficiary's voice directly to a resistant family member during a home visit. P10: ``So I will take the phone, then this recording is there, will make them listen, make them understand.'' This is a creative and internally logical idea. It is also the information deficit model in a new medium: an authoritative voice delivering health content to a resistant party. The training tool is being repurposed as a delivery mechanism for more explanation, which is exactly what training is meant to move ASHAs beyond.

Two other ASHAs described interest in using the chatbot for clinical information retrieval. Both alternate uses raise distinct deployment questions around network reliability, transparency with beneficiaries, and appropriate scope that would need to be resolved before any rollout.

Prior work with ASHAs has documented a tendency to defer to AI outputs uncritically, including incorrect ones~\cite{wadhwa2025designing,okolo2024if}. A system positioned as an authoritative voice that speaks to beneficiaries on the ASHA's behalf would likely accelerate that dynamic and diminish the ASHA's own communicative agency. The chatbot design probe reactions confirm that this use case is not hypothetical, it is what at least some ASHAs will attempt if the boundary between training environment and persuasion tool is not clearly established.

\subsubsection{Workload, Connectivity, and Data Costs}
The positive first-impression responses should also be read cautiously. Of the 20 participants, 19 reported awareness of \textit{ASHA Saheli}, an LLM-based WhatsApp chatbot that answers clinical questions~\cite{ramjee2025ashabot}. While ASHAs currently have many digital tools at their disposal, this is the only LLM-based tool that many have been exposed to. Yet none had used Saheli in the past month, and several appeared to confuse it with other digital tools. When asked to describe how they used it, some were uncertain which digital tool was being discussed. High awareness and low active use in the same population suggests a tool that was introduced successfully but has not been integrated into regular practice.

The likely explanation is not interface quality. When P18 was asked when she last used ASHA Saheli, she replied: ``Madam, right now there is so much load of work on us,'' and proceeded to describe the volume of digital reporting tasks now required of ASHAs alongside their field responsibilities. P12 named a more material constraint: ``We have this information, but we don't have enough money to give information to people and use WhatsApp.'' Several ASHAs, when offered the chance to ask questions at the end of the session, raised occupational frustrations---workload, irregular incentive payments, digital reporting burden---rather than anything about the study itself. This is important context for any new tool. An ASHA who has spent the morning completing digital forms, has home visits in the afternoon, and is uncertain when her next incentive payment will arrive is not positioned to explore an optional training application, regardless of its quality. As a result, enthusiasm for our chatbot design probe may not translate into active use.

\section{Discussion}

Across scripted calls, interviews, and a chatbot design probe with 20 ASHAs, we identified communication challenges involving health information, household relationships, and practical constraints. In calls, participants often repeated health information, dismissed concerns, offered unspecified help, or named methods with limited explanation. Other exchanges showed engagement with concerns and conditional offers of support. Interviews described strategies extending across repeated visits and established relationships, complementing the immediate responses elicited by the calls. Participants’ training recollections emphasized health content and repeated explanation, while national materials also address listening and negotiation. The chatbot elicited favorable initial reactions and proposed uses beyond rehearsal, while interview accounts and interaction disruptions highlighted workload, cost, and connectivity constraints. Together, these findings inform what communication training should support, how AI roleplay might contribute, and what conditions could limit its use.

Our findings suggest that responding directly to a beneficiary’s concern is an incomplete basis for judging communication quality. The calls and interview accounts included examples of responses that addressed concerns while introducing other tensions. P12 countered anxiety about old-age support by invoking daughters’ caregiving, reproducing a gendered expectation. In her interview, P6 described offering to cook during a woman’s recovery, addressing a practical barrier through additional labor of her own. These examples direct attention to whose preferences a response supports, what assumptions it makes, and who would bear the work of carrying it out.

Contraceptive autonomy provides a basis for examining these questions. It directs attention to whether a woman can make and act on an informed and voluntary decision, including accepting, postponing, switching, or declining a method \cite{senderowicz2020contraceptive}. The distinction matters when a woman’s preferences and those of her family differ: household agreement may coexist with pressure, and an unresolved disagreement does not establish that counseling has failed. In most simulated calls, the caller’s own preference was not explicitly elicited, making it difficult to assess how the conversation’s closure related to what she wanted. Communication support should therefore help ASHAs explore and clarify the woman’s preferences, including whether she wants further information, assistance, family involvement, or continued discussion.

Supporting these preferences can require articulation work: coordinating information, privacy, household relationships, and practical assistance when people’s interests and capacities differ \cite{schmidt1992taking}. Our findings highlight the tensions within that coordination. Family involvement may provide assistance while increasing pressure, and personal offers of help may resolve an immediate barrier while expanding the ASHA’s responsibilities. Articulation work must therefore be examined in terms of how it distributes voice, responsibility, and labor. Training may help workers recognize and navigate these tensions, but cannot supply missing recovery support or remove household coercion. Support for beneficiaries should not depend on workers absorbing further unpaid responsibilities.

\begin{figure}
    \centering
    \includegraphics[width=0.75\linewidth]{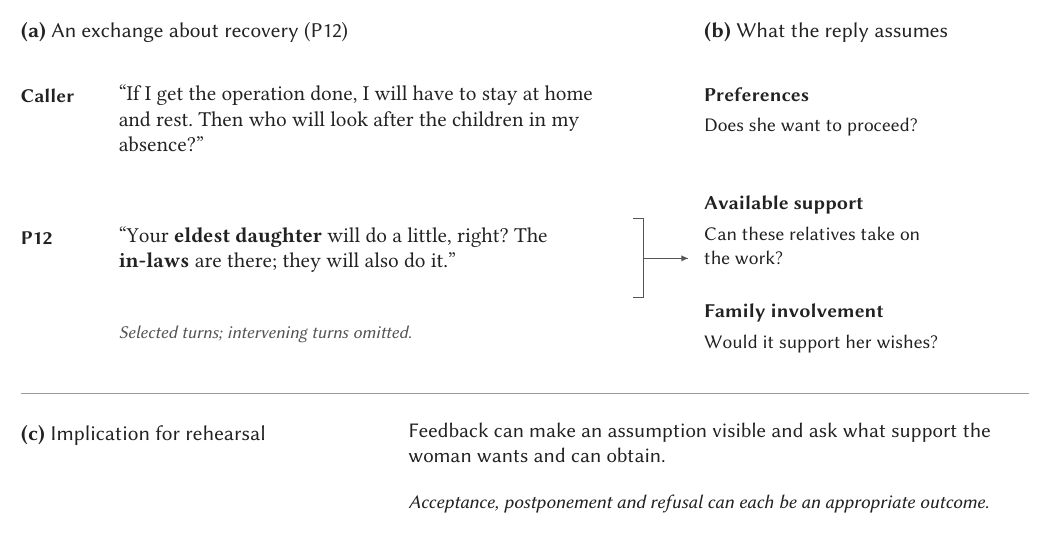}
    \caption{A reply can address a practical concern while presuming another person's willingness or capacity to help. The excerpt is from Section 4.3.2; emphasis is added. Panels (b) and (c) express the paper's interpretive and design argument (Section 5)}
    \Description{A three-part figure annotating a short exchange about sterilization
recovery. Panel (a), An exchange about recovery with P12, shows two quoted turns with
intervening turns omitted. The caller asks: "If I get the operation done, I will have
to stay at home and rest. Then who will look after the children in my absence?" P12
replies: "Your eldest daughter will do a little, right? The in-laws are there; they
will also do it," with "eldest daughter" and "in-laws" emphasised in bold. Panel (b),
What the reply assumes, connects P12's reply by a bracket to three unexamined
assumptions: preferences, whether she wants to proceed at all; available support,
whether these relatives can take on the work; and family involvement, whether it
would support her wishes. Panel (c), Implication for rehearsal, states that feedback
can make an assumption visible and ask what support the woman wants and can obtain,
and that acceptance, postponement and refusal can each be an appropriate outcome.}
    \label{fig:3}
\end{figure}

Figure~\ref{fig:3} illustrates these tensions through P12's response to a concern about household work during recovery. Her reply invokes potential helpers without establishing whether that support is available or wanted, illustrating the assumptions that AI-supported rehearsal could help trainees examine.

These tensions give AI-supported roleplay a specific focus for investigation. A system could reward a response for addressing the stated concern while overlooking an assumed preference, an unsupported promise, or additional labor assigned to someone else. Suchman’s account of situated action helps clarify the limits of prescribing an ideal reply: scripts can provide resources for action without determining what is appropriate in a particular encounter \cite{suchman1987plans}. Roleplay could instead make exchanges available for reflection, helping trainees examine what the caller expressed, what their response assumed, and what remains unknown. Feedback should leave room for asking permission, revising an offer, or accepting a decision to stop. Evaluation should examine whether these processes support women’s autonomy. Systems that reward agreement or completed uptake risk obscuring whether the woman’s preferences were respected and whose labor made that outcome possible \cite{star1999layers}.

\subsection{From Training Guidance to Supported Practice}

National materials include guidance on active listening, dialogue, negotiation, and difficult situations~\cite{nrhm2013asha,nhsrc_asha_induction_2021}. Participants’ recollections emphasized health content and repeated explanation, raising questions about how communication guidance is taught, practiced, and reinforced. Our data cannot distinguish the contributions of local delivery, supervision, recall, and working conditions.

MI-based training is one candidate worth testing because it specifies observable practices such as listening, eliciting preferences, asking permission, and supporting choice. It has also been taught to lay and community health workers in low-resource settings~\citep{mokhele2024improving}. Whether it can be successfully adapted for ASHAs remains an empirical question. A future program should provide locally adapted practice and compensation for participants’ time. Evaluation should examine communication processes and autonomy-related outcomes, including women’s own assessments of whether counseling was respectful and useful.

\subsection{Design Considerations for AI-Supported Roleplay Training}

These considerations draw on the three data sources differently. The interactional categories in Section 4.1 come directly from the call analysis. The boundary between rehearsal and client-facing use follows from probe reactions together with prior work on AI overreliance in this population. The delivery and compensation considerations draw on participants' interview accounts of workload and cost.

\subsubsection{Keep AI in a Rehearsal Role}

Participants did not always interpret the probe only as rehearsal. Two wanted to play its audio to family members, and two wanted to use it for clinical information. Prior work also describes cases in which ASHAs deferred to incorrect AI outputs~\cite{wadhwa2025designing,okolo2024if}.

A roleplay system should be presented as a practice environment. Generated audio should not be framed as approved material for home visits. Any client-facing use would need separate provisions for consent, disclosure, clinical validation, privacy, responsibility, and the woman's ability to refuse AI involvement. A system that supplies lines for delivery may reduce the ASHA's role to repeating generated advice. In rehearsal, she still has to decide what fits the situation, ideally with access to human supervision.

\subsubsection{Feedback Should Describe the Exchange}

The concern-evading taxonomy gives a feedback system something specific to comment on. It can note that the caller raised household labor and the trainee answered with pregnancy risks. This would ground feedback in an identifiable feature of the exchange as opposed to generic statements like ``be empathetic'' or ``listen better.''

Feedback should stay descriptive rather than verdict-like. Family-planning conversations rarely have one correct next turn, and a model cannot know the household, the woman's safety, or whether she wants to continue. For example:

\begin{quote}
The caller raised concern about pressure from her mother-in-law, but you returned to health risks. Before providing more information, consider acknowledging the concern and asking what kind of support she might like. She may also decide not to continue the discussion or not to adopt a method.
\end{quote}

This describes the exchange without presenting one required reply. Clinical content also needs separate handling. Statements about side effects, recovery, method eligibility, and birth spacing should be checked against approved sources rather than generated freely. When uncertain, the system should direct the trainee to approved guidance or a supervisor. 

\subsubsection{Support Adaptation Without Creating Another Script}

Roleplay should not replace one repeated argument with another script. Scenarios should vary whether the caller is asking for information, expressing ambivalence, making an informed refusal, describing household pressure, or raising a safety concern. Continued persuasion is not appropriate across these situations. An ambivalent caller may invite open inquiry and information with permission. A considered refusal may call for affirmation and an open door for future questions.

A system should not measure success through agreement, reduced hesitation, uptake, or speed. Candidate process measures include whether the trainee elicited the caller's preference, acknowledged the concern, asked permission before advising, provided accurate information, recognized refusal, and identified a safe next step when the caller wanted one. Some of these measures overlap with MI fidelity tools such as MITI~\citep{moyers2007revised}, which measure how well a practitioner is using MI and offer a method for providing structured, formal feedback for improvement. However, use in short Hindi-language simulations would require adaptation and validation.

\subsubsection{Account for ASHAs' Time and Access}

Notably, favorable first reactions to the design probe do not show sustained usefulness. Each participant encountered the probe once, with researchers present. Four sessions were disrupted by connectivity problems. Participants also described reporting burden, data costs, and irregular payments. These conditions will shape whether the system is used more than once.

A configuration for future evaluation we envision from these findings is accordingly minimal: a one- to two-week program in which ASHAs call a dedicated phone line each day, complete a five-minute simulated conversation with an AI beneficiary, receive immediate verbal feedback, and are paid a nominal amount for each completed session. The scenarios would be drawn directly from the case types in this study, birth spacing with family resistance, sterilization with son preference, Antara hesitancy, using resistance scripts grounded in the objection patterns we observed. There would be no app installation, no learning a new interface, and no requirement for data connectivity beyond a standard voice call.

Importantly, ASHAs should be compensated for each completed session. Participating in training competes with income-generating work, and the time cost is real. Compensation changes the framing: ASHAs are engaging in valued professional development, not using a free tool in their spare time.  This recommendation also exposes a gap in our own study: participants were not compensated for their session time. Guidance from our partner non-profit shaped the decision in this context.

\subsection{Limitations}
Our study has several limitations. First, the sample included 20 ASHAs recruited through one partner-linked pool in one district of Rajasthan. The findings are not prevalence estimates and should not be generalized to ASHAs nationally, to other health topics, or to regions with different training arrangements. Second, the study used scripted telephone calls enacted by one actor. The calls lacked the relationship history, household presence, privacy conditions, and repeated encounters of in-person work. The unannounced format, partner and researcher presence, perceived evaluation, and the actor's delivery may have affected responses. Interviews were shaped by different conditions, including recall and self-presentation. Neither source provides direct access to routine practice. Third, the study involved deception: at the guidance of the local partner, participants initially understood the scripted caller to be a beneficiary seeking guidance, and the call was recorded before disclosure and consent for research use. The full protocol, including these procedures, received institutional ethics approval. Participants could decline participation and research use of the recording after the call, in which case the recording would be deleted. However, this opportunity could not undo the initial interaction or any discomfort it generated. Recruitment through an organization with an ongoing relationship with participants may also have made refusal feel difficult, particularly given indications that some participants perceived the study as a performance evaluation. In line with the partner’s guidance, participants received no direct compensation for study participation, but did receive data packs through a separate deployment run by the partner. While partner guidance informed this decision, it did not remove participants' time costs or our responsibility to consider them. 
Fourth, the calls staged two family-planning scenarios. Women considering family-planning services, family members, supervisors, clinicians, and reproductive-rights practitioners were not participants. We cannot determine how women would interpret the responses or which outcomes they would find respectful and useful. Finally, the chatbot was a brief design probe rather than a tested intervention. Participants encountered it once, researchers were present, and four sessions were interrupted. We did not assess repeated use, learning, client outcomes, clinical safety, or the reliability of automated Hindi-language feedback.

\section{Conclusion}

We studied 20 ASHAs in Rajasthan through scripted calls, interviews, and a chatbot design probe. In the calls, participants often returned to health-risk or equity arguments after the caller raised a social or material concern. Other responses contradicted a concern, promised help without a plan, or named methods with little explanation. Because these were short simulations conducted under observation, we do not treat them as a direct record of routine practice.

A smaller number of responses acknowledged the caller's concern, asked permission before involving family members, or kept open the possibility of further contact. Motivational Interviewing provides vocabulary for interpreting these moves, while contraceptive autonomy scholarship changes the standard against which they should be judged. Good counseling does not require uptake. A woman may understand the options and still postpone or decline.

AI-supported roleplay could help trainees examine how they responded to a particular concern. It should not supply authoritative scripts, score success through agreement or uptake, or act as a persuasive voice during home visits. Whether such feedback can be delivered accurately in Hindi, used within ASHAs' working conditions, and, most importantly, lead to improved communication and better healthcare outcomes remains to be tested.

\section*{Acknowledgments}
We thank Surya Vaishnav for acting in the simulated call scenarios, and Mahendra Kumar Meena and Adil Khan for coordinating the field interviews with ASHA participants. This project was funded in part by the University of Pennsylvania’s Penn Development Research Initiative - DevLab (NKRS), Center for the Advanced Study of INDIA (NKRS), and Penn Global (NKRS). The funders had no role in the design and conduct of the study; collection, management, analysis, and interpretation of the data; preparation, review, or approval of the manuscript; and decision to submit the manuscript for publication.
We used LLMs (e.g. ChatGPT, Claude) to support drafting (e.g. paraphrasing for clarity) and formatting during manuscript preparation. All generated text was reviewed and verified by the authors. All ideas and interpretations presented in this paper are solely those of the authors.

\section*{Ethics and Privacy Statement}

The protocol, including the unannounced simulated call, delayed disclosure, and post-interaction consent for retaining the recording, was reviewed and approved by the Institutional Review Board at [Institution]. Publication of this work carries risks of normalizing deceptive data collection with frontline workers and of encouraging AI roleplay systems to be repurposed as persuasive authorities in family-planning encounters, where they could disclose private household information, provide inaccurate clinical advice, or undermine contraceptive autonomy. We therefore limit our claims to responses elicited under the study conditions and argue that future systems should remain rehearsal tools, require separate consent and clinical review for any client-facing use, protect ASHAs' and clients' privacy, and treat informed refusal as a legitimate outcome.

\bibliographystyle{ACM-Reference-Format}
\bibliography{sample-base}

@String{Computing = "Computing" }

@String{Computer = "{IEEE} Computer" }

@String{Academic = "Academic Press" }

@String{Springer = "Springer-Verlag" }

@article{braun2006using,
  title={Using thematic analysis in psychology},
  author={Braun, Virginia and Clarke, Victoria},
  journal={Qualitative research in psychology},
  volume={3},
  number={2},
  pages={77--101},
  year={2006},
  publisher={Taylor \& Francis}
}

@article{CHAWLA2025100134,
title = {Performance and challenges of Accredited Social Health Activists (ASHAs) in delivering key Maternal and Newborn Health (MNH) services in India: A systematic review and meta-analyses},
journal = {SSM - Health Systems},
volume = {5},
pages = {100134},
year = {2025},
issn = {2949-8562},
doi = {https://doi.org/10.1016/j.ssmhs.2025.100134},
url = {https://www.sciencedirect.com/science/article/pii/S2949856225000868},
author = {Sukriti Chawla and Chandan Kumar and Montu Bose and Shikha M. Shrivastav},
}

@article{shrivastava2016measuring,
  title={Measuring communication competence and effectiveness of ASHAs (accredited social health activist) in their leadership role at rural settings of Uttar Pradesh (India)},
  author={Shrivastava, Archana and Srivastava, Arun},
  journal={Leadership in Health Services},
  volume={29},
  number={1},
  pages={69--81},
  year={2016},
  publisher={Emerald Group Publishing Limited}
}

@article{scott2022we,
  title={’[We] learned how to speak with love’: a qualitative exploration of accredited social health activist (ASHA) community health worker experiences of the Mobile Academy refresher training in Rajasthan, India},
  author={Scott, Kerry and Ummer, Osama and Chamberlain, Sara and Sharma, Manjula and Gharai, Dipanwita and Mishra, Bibha and Choudhury, Namrata and LeFevre, Amnesty Elizabeth},
  journal={BMJ open},
  volume={12},
  number={6},
  pages={e050363},
  year={2022},
  publisher={British Medical Journal Publishing Group}
}

@article{goel2019effectiveness,
  title={Effectiveness of a quality improvement program using Difference-in-Difference analysis for home based newborn care--results of a community intervention trial},
  author={Goel, Akhil Dhanesh and Gosain, Mudita and Amarchand, Ritvik and Sharma, Hanspria and Rai, Sanjay and Kapoor, Suresh K and Krishnan, Anand},
  journal={The Indian Journal of Pediatrics},
  volume={86},
  number={11},
  pages={1028--1035},
  year={2019},
  publisher={Springer}
}

@article{okolo2024if,
  title={" If it is easy to understand then it will have value": Examining Perceptions of Explainable AI with Community Health Workers in Rural India},
  author={Okolo, Chinasa T and Agarwal, Dhruv and Dell, Nicola and Vashistha, Aditya},
  journal={Proceedings of the ACM on Human-Computer Interaction},
  volume={8},
  number={CSCW1},
  pages={1--28},
  year={2024},
  publisher={ACM New York, NY, USA}
}

@inproceedings{okolo2021cannot,
  title={“It cannot do all of my work”: community health worker perceptions of AI-enabled mobile health applications in rural India},
  author={Okolo, Chinasa T and Kamath, Srujana and Dell, Nicola and Vashistha, Aditya},
  booktitle={Proceedings of the 2021 CHI conference on human factors in computing systems},
  pages={1--20},
  year={2021}
}

@article{smartkc,
author = {Gairola, Siddhartha and Bohra, Murtuza and Shaheer, Nadeem and Jayaprakash, Navya and Joshi, Pallavi and Balasubramaniam, Anand and Murali, Kaushik and Kwatra, Nipun and Jain, Mohit},
title = {SmartKC: Smartphone-based Corneal Topographer for Keratoconus Detection},
year = {2022},
issue_date = {Dec 2021},
publisher = {Association for Computing Machinery},
address = {New York, NY, USA},
volume = {5},
number = {4},
url = {https://doi.org/10.1145/3494982},
doi = {10.1145/3494982},
journal = {Proc. ACM Interact. Mob. Wearable Ubiquitous Technol.},
month = dec,
articleno = {155},
numpages = {27}
}

@inproceedings{ramjee2025ashabot,
  title={Ashabot: An llm-powered chatbot to support the informational needs of community health workers},
  author={Ramjee, Pragnya and Chhokar, Mehak and Sachdeva, Bhuvan and Meena, Mahendra and Abdullah, Hamid and Vashistha, Aditya and Nagar, Ruchit and Jain, Mohit},
  booktitle={Proceedings of the 2025 CHI Conference on Human Factors in Computing Systems},
  pages={1--22},
  year={2025}
}

@article{wadhwa2025designing,
  title={Designing with Culture: How Social Norms Shape Trust and Preference in Health Chatbots},
  author={Wadhwa, Arpita and Vashistha, Aditya and Jain, Mohit},
  journal={arXiv preprint arXiv:2509.15575},
  year={2025}
}

@inproceedings{ming2022invisible,
  title={Invisible work in two frontline health contexts},
  author={Ming, Joy and Kamath, Srujana and Kuo, Elizabeth and Sterling, Madeline and Dell, Nicola and Vashistha, Aditya},
  booktitle={Proceedings of the 5th ACM SIGCAS/SIGCHI Conference on Computing and Sustainable Societies},
  pages={139--151},
  year={2022}
}

@inproceedings{solano2024explorable,
  title={Explorable Explainable AI: Improving AI Understanding for Community Health Workers in India},
  author={Solano-Kamaiko, Ian Ren{\'e} and Mishra, Dibyendu and Dell, Nicola and Vashistha, Aditya},
  booktitle={Proceedings of the 2024 CHI Conference on Human Factors in Computing Systems},
  pages={1--21},
  year={2024}
}

@article{xiao2025use,
  title={Is the use of standardized patients more effective than role-playing in medical education? A meta-analysis},
  author={Xiao, Jingyuan and Fu, Xinjian},
  journal={Frontiers in Medicine},
  volume={12},
  pages={1601116},
  year={2025},
  publisher={Frontiers Media SA}
}

@article{ross2024impact,
  title={The impact of standardized patients on first-year nursing students’ communication skills},
  author={Ross, Jennifer Gunberg and Furman, Gail and Scheve, Ann},
  journal={Clinical simulation in nursing},
  volume={89},
  pages={101513},
  year={2024},
  publisher={Elsevier}
}

@inproceedings{sehgal2025pal,
  title={PAL: Designing Conversational Agents as Scalable, Cooperative Patient Simulators for Palliative-Care Training},
  author={Sehgal, Neil KR and Kambhamettu, Hita and Chang, Allen and Zhu, Andrew and Ungar, Lyle and Guntuku, Sharath Chandra},
  booktitle={Companion Publication of the 2025 Conference on Computer-Supported Cooperative Work and Social Computing},
  pages={346--350},
  year={2025}
}

@article{baseman2025poker,
  title={'Poker with Play Money': Exploring Psychotherapist Training with Virtual Patients},
  author={Baseman, Cynthia M and Hasan, Masum and Swinger, Nathaniel and Rauch, Sheila AM and Hoque, Ehsan and Arriaga, Rosa I},
  journal={Proceedings of the ACM on Human-Computer Interaction},
  volume={9},
  number={7},
  pages={1--29},
  year={2025},
  publisher={ACM New York, NY, USA}
}

@article{haut2025ai,
  title={Ai standardized patient improves human conversations in advanced cancer care},
  author={Haut, Kurtis and Hasan, Masum and Carroll, Thomas and Epstein, Ronald and Sen, Taylan and Hoque, Ehsan},
  journal={arXiv preprint arXiv:2505.02694},
  year={2025}
}

@article{goldwater2025community,
  title={Community health workers’ counseling is based on a deficit model of behavior change},
  author={Goldwater, Micah B and Hashmi, Faiz A and Mondal, Sudipta and Legare, Cristine H},
  journal={PLOS Global Public Health},
  volume={5},
  number={7},
  pages={e0004167},
  year={2025},
  publisher={Public Library of Science San Francisco, CA USA}
}

@book{nhsrc_asha_induction_2021,
  title        = {Induction Training Module for ASHAs},
  author       = {{Ministry of Health and Family Welfare, Government of India}},
  year         = {2021},
  institution  = {National Health Systems Resource Centre (NHSRC)},
  address      = {New Delhi, India},
  url          = {https://nhsrcindia.org/sites/default/files/2021-02/Induction-Training-Module-for-ASHA-English_0.pdf},
  note         = {Training manual for Accredited Social Health Activists (ASHAs)}
}

@incollection{wang2025dismantling,
  title={Dismantling the deficit model of science communication using Ludwik Fleck’s theory of thinking collectives},
  author={Wang, Victoria Min-Yi},
  booktitle={Values, Pluralism, and Pragmatism: Themes from the Work of Matthew J. Brown},
  pages={117--137},
  year={2025},
  publisher={Springer}
}

@article{kelly2016changing,
  title={Why is changing health-related behaviour so difficult?},
  author={Kelly, Michael P and Barker, Mary},
  journal={Public health},
  volume={136},
  pages={109--116},
  year={2016},
  publisher={Elsevier}
}

@article{seethaler2019science,
  title={Science, values, and science communication: Competencies for pushing beyond the deficit model},
  author={Seethaler, Sherry and Evans, John H and Gere, Cathy and Rajagopalan, Ramya M},
  journal={Science Communication},
  volume={41},
  number={3},
  pages={378--388},
  year={2019},
  publisher={SAGE Publications Sage CA: Los Angeles, CA}
}

@article{NewlinCanzone2013,
  author = {Newlin-Canzone, ET and Scerbo, MW and Gliva-McConvey, G and Wallace, AM},
  title = {The cognitive demands of standardized patients: understanding limitations in attention and working memory with the decoding of nonverbal behavior during improvisations},
  journal = {Simulation in Healthcare},
  year = {2013},
  volume = {8},
  number = {4},
  pages = {207--214},
}

@article{Hart2016,
  author = {Hart, JA and Chilcote, DR},
  title = {"Won't You Be My Patient?": Preparing Theater Students as Standardized Patients},
  journal = {Journal of Nursing Education},
  year = {2016},
  volume = {55},
  number = {3},
  pages = {168--171},
}

@article{Brenner2009,
  author = {Brenner, AM},
  title = {Uses and limitations of simulated patients in psychiatric education},
  journal = {Academic Psychiatry},
  year = {2009},
  volume = {33},
  number = {2},
  pages = {112--119},
}

@article{Nagpal2021,
  author = {Nagpal, V and Philbin, M and Yazdani, M and Veerreddy, P and Fish, D and Reidy, J},
  title = {Effective Goals-of-Care Conversations: From Skills Training to Bedside},
  journal = {MedEdPORTAL},
  year = {2021},
  month = {Mar},
  volume = {17},
  pages = {11122},
  pmid = {33768153},
  pmcid = {PMC7970639}
}

@article{ray2024experiences,
  title={Experiences of “Antara”: The Injectable Contraceptive in Rural Indian Women Presenting to a Tertiary Care Hospital of Eastern India: S. Ray et al.},
  author={Ray, Sabyasachi and Sing, Kinkar and Biswas, Titol and Manepalli, Siva Tejaswi and Chaturvedi, Akanksha},
  journal={The Journal of Obstetrics and Gynecology of India},
  volume={74},
  number={3},
  pages={243--249},
  year={2024},
  publisher={Springer}
}

@article{agrawal2022insights,
  title={Insights from client experience with injection medroxyprogesterone acetate (MPA) in India: lessons from the field},
  author={Agrawal, Swati and Sood, Shilpa and Chopra, Kanika and Singh, Anuradha and Gupta, Aparajita and Singh, Shalini and Puri, Manju},
  journal={The Journal of Obstetrics and Gynecology of India},
  volume={72},
  number={Suppl 1},
  pages={262--266},
  year={2022},
  publisher={Springer}
}

@article{oliveira2014dominance,
  title={Dominance of sterilization and alternative choices of contraception in India: an appraisal of the socioeconomic impact},
  author={Oliveira, Isabel Tiago de and Dias, Jos{\'e} G and Padmadas, Sabu S},
  journal={PLoS One},
  volume={9},
  number={1},
  pages={e86654},
  year={2014},
  publisher={Public Library of Science San Francisco, USA}
}

@article{schoeman2019digital,
  title={Digital tools for training frontline health workers in low and middle-income countries: A systematic review},
  author={Schoeman, Fransien},
  year={2019}
}

@article{long2018digital,
  title={Digital technologies for health workforce development in low-and middle-income countries: a scoping review},
  author={Long, Lesley-Anne and Pariyo, George and Kallander, Karin},
  journal={Global Health: Science and Practice},
  volume={6},
  number={Supplement 1},
  pages={S41--S48},
  year={2018},
  publisher={Global Health: Science and Practice}
}

@article{star1999layers,
  title={Layers of silence, arenas of voice: The ecology of visible and invisible work},
  author={Star, Susan Leigh and Strauss, Anselm},
  journal={Computer supported cooperative work (CSCW)},
  volume={8},
  number={1},
  pages={9--30},
  year={1999},
  publisher={Springer}
}

@article{schmidt1992taking,
  title={Taking CSCW seriously: Supporting articulation work},
  author={Schmidt, Kjeld and Bannon, Liam},
  journal={Computer supported cooperative work (CSCW)},
  volume={1},
  number={1},
  pages={7--40},
  year={1992},
  publisher={Springer}
}

@article{verdezoto2021invisible,
  title={The invisible work of maintenance in community health: challenges and opportunities for digital health to support frontline health workers in Karnataka, South India},
  author={Verdezoto, Nervo and Bagalkot, Naveen and Akbar, Syeda Zainab and Sharma, Swati and Mackintosh, Nicola and Harrington, Deirdre and Griffiths, Paula},
  journal={Proceedings of the ACM on Human-Computer Interaction},
  volume={5},
  number={CSCW1},
  pages={1--31},
  year={2021},
  publisher={ACM New York, NY, USA}
}

@book{suchman1987plans,
  title={Plans and situated actions: The problem of human-machine communication},
  author={Suchman, Lucille Alice},
  year={1987},
  publisher={Cambridge university press}
}

@book{miller2012motivational,
  title={Motivational interviewing: Helping people change},
  author={Miller, William R and Rollnick, Stephen},
  year={2012},
  publisher={Guilford press}
}

@article{mokhele2024improving,
  title={Improving patient-centred counselling skills among lay healthcare workers in South Africa using the Thusa-Thuso motivational interviewing training and support program},
  author={Mokhele, Idah and Sineke, Tembeka and Vujovic, Marnie and Ruiter, Robert AC and Miot, Jacqui and Onoya, Dorina},
  journal={PLOS global public health},
  volume={4},
  number={4},
  pages={e0002611},
  year={2024},
  publisher={Public Library of Science San Francisco, CA USA}
}

@manual{nrhm2013asha,
  author       = {{National Rural Health Mission}},
  title        = {Induction Training Module for {ASHAs}: A Consolidated Version of Modules 1 to 5 for Newly Selected {ASHAs}},
  organization = {{Ministry of Health and Family Welfare, Government of India}},
  year         = {2013},
  url          = {https://nhm.gov.in/images/pdf/communitisation/asha/ASHA_Induction_Module_English.pdf},
  urldate      = {2026-08-05}
}

@inproceedings{dell2012yours,
  title={" Yours is better!" participant response bias in HCI},
  author={Dell, Nicola and Vaidyanathan, Vidya and Medhi, Indrani and Cutrell, Edward and Thies, William},
  booktitle={Proceedings of the sigchi conference on human factors in computing systems},
  pages={1321--1330},
  year={2012}
}

@article{bowen2006grounded,
  title={Grounded theory and sensitizing concepts},
  author={Bowen, Glenn A},
  journal={International journal of qualitative methods},
  volume={5},
  number={3},
  pages={12--23},
  year={2006},
  publisher={SAGE Publications Sage CA: Los Angeles, CA}
}

@article{allotey2025using,
  title={Using of Mystery Clients’ Experiences and Feedback to Improve the Quality of Family Planning Services in Northern Ghana: Evidence From an Uncontrolled Quasi-experimental Study},
  author={Allotey, Naa-Korkor and Appiah, Evans K and Torpey, Kwasi},
  journal={Cureus},
  volume={17},
  number={8},
  year={2025},
  publisher={Cureus}
}

@article{das2016impact,
  title={The impact of training informal health care providers in India: A randomized controlled trial},
  author={Das, Jishnu and Chowdhury, Abhijit and Hussam, Reshmaan and Banerjee, Abhijit V},
  journal={Science},
  volume={354},
  number={6308},
  pages={aaf7384},
  year={2016},
  publisher={American Association for the Advancement of Science}
}

@article{cox2023toolkit,
  title={What is in the toolkit (and what are the tools)? How to approach the study of doctor--patient communication},
  author={Cox, Caitr{\'\i}ona and Fritz, Zo{\"e}},
  journal={Postgraduate medical journal},
  volume={99},
  number={1172},
  pages={631--638},
  year={2023},
  publisher={Oxford University Press}
}

@article{moyers2007revised,
  title={Revised global scales: Motivational interviewing treatment integrity 3.0 (MITI 3.0)},
  author={Moyers, TB and Martin, T and Manuel, JK and Miller, WR and Ernst, D},
  journal={University of New Mexico, Center on Alcoholism, Substance Abuse and Addictions (CASAA)},
  volume={28},
  year={2007}
}

@article{gwatkin1979political,
  author  = {Gwatkin, Davidson R.},
  title   = {Political Will and Family Planning: The Implications of India's Emergency Experience},
  journal = {Population and Development Review},
  year    = {1979},
  volume  = {5},
  number  = {1},
  pages   = {29--59},
  doi     = {10.2307/1972317}
}

@article{williams2014storming,
  author  = {Williams, Rebecca J.},
  title   = {Storming the Citadels of Poverty: Family Planning under the Emergency in India, 1975--1977},
  journal = {The Journal of Asian Studies},
  year    = {2014},
  volume  = {73},
  number  = {2},
  pages   = {471--492},
  doi     = {10.1017/S0021911813002350}
}

@manual{nhm2025ashaincentives,
  author       = {{National Health Mission}},
  title        = {{ASHA} Incentives},
  organization = {Ministry of Health and Family Welfare, Government of India},
  year         = {2025},
  month        = {July},
  note         = {See Family Planning incentive schedule}
}

@techreport{iips2026nfhs6,
  author      = {{International Institute for Population Sciences (IIPS)}},
  title       = {National Family Health Survey (NFHS-6), 2023--24: India and State/UT Fact Sheets},
  institution = {Ministry of Health and Family Welfare, Government of India},
  address     = {Mumbai},
  year        = {2026}
}

@article{bruce1990fundamental,
  author  = {Bruce, Judith},
  title   = {Fundamental Elements of the Quality of Care: A Simple Framework},
  journal = {Studies in Family Planning},
  year    = {1990},
  volume  = {21},
  number  = {2},
  pages   = {61--91},
  doi     = {10.2307/1966669}
}

@article{senderowicz2020contraceptive,
  author  = {Senderowicz, Leigh},
  title   = {Contraceptive Autonomy: Conceptions and Measurement of a Novel Family Planning Indicator},
  journal = {Studies in Family Planning},
  year    = {2020},
  volume  = {51},
  number  = {2},
  pages   = {161--176},
  doi     = {10.1111/sifp.12114}
}

@article{senderowicz2019obligated,
  author  = {Senderowicz, Leigh},
  title   = {{``I Was Obligated to Accept'': A Qualitative Exploration of Contraceptive Coercion}},
  journal = {Social Science \& Medicine},
  year    = {2019},
  volume  = {239},
  pages   = {112531},
  doi     = {10.1016/j.socscimed.2019.112531}
}

@article{holt2021gujarat,
  author  = {Holt, Kelsey and Uttekar, Bella Vasant and Reed, Reiley and Adams, Madeline and Kanchan, Lakhwani and Langer, Ana and Barge, Sandhya},
  title   = {Understanding Quality of Contraceptive Services from Women's Perspectives in Gujarat, India: A Focus Group Study},
  journal = {BMJ Open},
  year    = {2021},
  volume  = {11},
  number  = {10},
  pages   = {e049260},
  doi     = {10.1136/bmjopen-2021-049260}
}

@article{aruldas2017care,
  title={Care-seeking behaviors for maternal and newborn illnesses among self-help group households in Uttar Pradesh, India},
  author={Aruldas, Kumudha and Kant, Aastha and Mohanan, PS},
  journal={Journal of Health, Population and Nutrition},
  volume={36},
  number={Suppl 1},
  pages={49},
  year={2017},
  publisher={Springer}
}

@inproceedings{steenstra2025scaffolding,
  title={Scaffolding empathy: Training counselors with simulated patients and utterance-level performance visualizations},
  author={Steenstra, Ian and Nouraei, Farnaz and Bickmore, Timothy},
  booktitle={Proceedings of the 2025 chi conference on human factors in computing systems},
  pages={1--22},
  year={2025}
}

@inproceedings{ramachandran2010mobile,
  title={Mobile-izing health workers in rural India},
  author={Ramachandran, Divya and Canny, John and Das, Prabhu Dutta and Cutrell, Edward},
  booktitle={Proceedings of the SIGCHI conference on human factors in computing systems},
  pages={1889--1898},
  year={2010}
}

@inproceedings{ismail2021ai,
  author    = {Ismail, Azra and Kumar, Neha},
  title     = {{AI} in Global Health: The View from the Front Lines},
  booktitle = {Proceedings of the 2021 CHI Conference on Human Factors in Computing Systems},
  year      = {2021},
  publisher = {Association for Computing Machinery},
  numpages  = {21},
  doi       = {10.1145/3411764.3445130}
}

@inproceedings{yadav2021illustrating,
  author    = {Yadav, Deepika and Malik, Prerna and Dabas, Kirti
               and Singh, Pushpendra},
  title     = {Illustrating the Gaps and Needs in the Training Support
               of Community Health Workers in India},
  booktitle = {Proceedings of the 2021 CHI Conference on Human Factors in Computing Systems},
  year      = {2021},
  publisher = {Association for Computing Machinery},
  numpages  = {16},
  doi       = {10.1145/3411764.3445111}
}

@inproceedings{sehgal2025exploring,
  title={Exploring Socio-Cultural challenges and opportunities in designing mental health chatbots for adolescents in India},
  author={Sehgal, Neil KR and Kambhamettu, Hita and Matam, Sai Preethi and Ungar, Lyle and Guntuku, Sharath Chandra},
  booktitle={Proceedings of the Extended Abstracts of the CHI Conference on Human Factors in Computing Systems},
  pages={1--7},
  year={2025}
}

\appendix

\newpage
\section{APPENDIX}

\begin{table}[h]

\centering
\caption{Years of experience for the 20 ASHA participants included in the study.}
\label{tab:participant-experience}
\begin{tabular}{| l  |l |l|}
\hline
\textbf{Participant ID} & \textbf{Years of Experience}  &\textbf{Call Scenario}\\
\hline
P1 & 8   &Sterilization / Son preference\\
\hline
P2 & 14   &Sterilization / Son preference\\
\hline
P3 & 10  &Sterilization / Son preference\\
\hline
P4 & 14  &Antara / Birth Spacing\\
\hline
P5 & 19  &Sterilization / Son preference\\
\hline
P6 & 10   &Sterilization / Son preference\\
\hline
P7 & 19  &Antara / Birth Spacing\\
\hline
P8 & 21  &Antara / Birth Spacing\\
\hline
P9 & 8  &Antara / Birth Spacing\\
\hline
P10 & 12  &Antara / Birth Spacing\\
\hline
P11 & 13  &Antara / Birth Spacing\\
\hline
P12 & 21  &Sterilization / Son preference\\
\hline
P13 & 19  &Sterilization / Son preference\\
\hline
P14 & 12  &Sterilization / Son preference\\
\hline
P15 & 11  &Sterilization / Son preference\\
\hline
P16 & 10  &Sterilization / Son preference\\
\hline
P17 & 10  &Antara / Birth Spacing\\
\hline
P18 & 11  &Antara / Birth Spacing\\
\hline
P19 & 19  &Antara / Birth Spacing\\
\hline
P20 & 7  &Antara / Birth Spacing\\
\hline

\end{tabular}

\end{table}

\end{document}